\documentclass{PoS}

\usepackage{amsmath,amssymb,amsthm,latexsym}
\usepackage{stmaryrd,wasysym,upgreek,mathrsfs,dsfont}

\newtheorem{theorem}{Theorem}[section]
\newcommand{\be}{\begin{equation}}
\newcommand{\ee}{\end{equation}}
\newcommand{\bea}{\begin{eqnarray}}
\newcommand{\eea}{\end{eqnarray}}

\newcommand{\prf}{\noindent{\bf Proof}\ }

\newcommand{\Tr}{{\rm Tr}}

\newcommand{\cV}{{\cal{V}}}

\newcommand{\cT}{{\cal{T}}}

\title{Construction of  $2$-dimensional Grosse-Wulkenhaar Model}

\ShortTitle{Constructive GW2}

\author{\speaker{Zhituo Wang}\\
        Laboratoire de Physique Th\'eorique, CNRS UMR 8627,\\
Universit\'e Paris XI,  F-91405 Orsay Cedex, France\\
        E-mail: \email{ zhituo.wang@th.u-psud.fr}}

\abstract{In this talk we briefly report the recent work on the construction of
the 2-dimensional Grosse-Wulkenhaar model with the method of loop vertex
expansion \cite{Wanggw2}. We treat renormalization with this new tool, adapt Nelson's
argument and prove Borel summability of the perturbation series.
This is the first non-commutative quantum field theory
model to be built in a non-perturbative sense.}

\FullConference{Proceedings of the Corfu Summer Institute 2011 School and Workshops on Elementary Particle Physics and Gravity\\
		September 4-18, 2011\\
		Corfu, Greece}

\begin{document}

\section{Introduction}
It is believed that in quantum field theory including gravity, space-time would become
fuzzy at distances compare to the Planck scale. Noncommutative quantum field theory (NCQFT) \cite{Douglas,Szabo}  is a possible way
to study the geometrical fuzzy structure and physics models in such fuzzy background. The simplest noncommutative space is the
Moyal space $\mathbb {R}^{D}_{\theta}$ with the coordinates obeying the commutation relation $[x_i,x_j]=i\theta_{ij}$, and the
simplest noncommutative quantum field theory model is a scale field theory with $\lambda\phi^4$
interaction defined on  $\mathbb {R}^{4}_{\theta}$. However such model suffers from the UV/IR mixing, namely,
the correlation function is still infrared divergent when integrating out the ultraviolet modes for certain non-planar
graphs \cite{Minwala}. Several years ago H. Grosse and R. Wulkenhaar solved this problem by introducing
a harmonic oscillator term in the propagator so that the theory fully obeys a new symmetry
called the Langmann-Szabo duality. They have proved
in a series of papers  \cite{GWp,GW0,GW1} (see also \cite{RVW}) that the noncommutative
$\phi^{\star 4}_4$ field theory possessing the Langmann-Szabo duality (which we
call the $GW_4$ model hereafter) is perturbatively renormalizable to all
orders.

After the work of Grosse and Wulkenhaar many other QFT models on Moyal space \cite{LsSZ2,LaSZ3,Fabien1,GrStei,GS1,GS2,GourWW}
or degenerate Moyal space  \cite{Wang1, GF} have also been shown to be {\emph{perturbatively}} renormalizable. More details could be found in
\cite{Bourbaphy, Raimar, Fabien}.

The $GW_4$ model is not only perturbatively renormalizable but also
asymptotically safe due to the vanishing of the $\beta$ function in the
ultraviolet regime \cite{beta1}, \cite{MagRiv}, \cite{DGMR}, which
means that there are neither Landau ghost which appears in commutative $\phi^4_4$ and QED
nor confinement for non-Abelian gauge theory.

So that $GW_4$  is a prime natural candidate for a full construction
of a four dimensional just  renormalizable quantum field theory \cite{Riv}.

In this paper we shall construct the $2$ dimensional Grosse-Wulkenhaar model ($GW_2$ for simplicity),
namely we prove that the perturbation series of the connected Schwinger's function is Borel summable,
as a warm-up towards building non-perturbatively the full $GW_4$ model.
The method we use is the Loop Vertex Expansion (LVE) invented by Vincent Rivasseau \cite{Riv1}
(see also \cite{Wanggw2,Riv1,RivMag,RivW1,RivW2,RWang} for more details),
which is a combination of the intermediate field techniques with the BKAR forest formula \cite{BK,AR1}.

In order to simplify the reading of this paper we briefly summerize the ideas and methods of the construction procedure.
We shall first obtain the intermediate field representation of the partition function for the GW2 model
by introducing the intermediate matrix field $\sigma$ and integrating out the original scalar field $\phi$. Then we use the
BKAR tree formula to derive the perturbation series of the connected Schwinger's function. However the amplitude is divergent due to the presence of the tadpoles. So we introduce a new expansion, called the cleaning expansion, to compensate the divergences. We should stop the cleaning
expansion when we have obtained enough convergent factor, as otherwise we would generate big combinatorial factors which diverge again. This is the analogue of the Nelson's argument for traditional constructive field theory \cite{Riv}.
After that we shall re-summing the uncompensated tadpoles. In the re-summing procedure we should analyze carefully the scales
of the indices such that not all tadpoles should be re-summed. This analysis plays a role of the traditional cluster expansion.
Then in the end we get the bound for the perturbation series and prove the Borel summability.

Another approachs towards the construction of $GW_4$, based on the combination of a Ward identity and
Schwinger-Dyson equations, is given in \cite{GW2}, and a numerical study of NCQFT models is given in \cite{num}.

\medskip
\noindent{\bf Acknowledgments}
The author is very grateful to his Phd supervisor Prof. Vincent Rivasseau for many very helpful discussions, as well as
the organizers of the Corfu 2011 Workshops on Elementary Particle Physics and Gravity for their invitation.

\section{Moyal space and Grosse-Wulkenhaar Model}
\subsection{The Moyal space}
The $D$-dimensional Moyal space $\mathbb {R}^{D}_{\theta}$ for $D$ even is generated by the
non-commutative coordinates $x^{\mu}$ that obey
the commutation relation $[ x^{\mu},x^{\nu}]=i\Theta^{\mu\nu}$, where
$\Theta$ is a $D\times D$ non-degenerate skew-symmetric
matrix. (see
\cite{GPW, Bourbaphy} for more details).

The 2-dimensional Grosse-Wulkenhaar model in the matrix basis is defined by (see \cite{Wanggw2} for more details):
\begin{eqnarray}
&&S[\phi] = 2\pi\theta \Tr \Big[ \frac{1}{2}
\phi \Delta\phi +: \frac{\lambda}{4}
\phi^4_\star:\Big]\nonumber\\&=&2\pi\theta \sum_{m,n,k,l} \Big[ \frac{1}{2}
\phi_{mn} \Delta_{mn;kl} \phi_{kl} + \frac{\lambda}{4}
:\phi_{mn} \phi_{nk} \phi_{kl} \phi_{lm}:\Big].
\end{eqnarray}
where $\phi_{mn}$ is a Hermitian matrix and
we have put the ultraviolet cutoff $\Lambda$ to the matrix indices. Namely, $m,\ n=1,\cdots, \Lambda$.
We have taken the wick ordering to the interaction term whose explicit form reads:
\begin{equation}\label{Wick1}
:\phi_{mn} \phi_{nk} \phi_{kl} \phi_{lm}:=\phi_{mn} \phi_{nk} \phi_{kl} \phi_{lm}-8\phi_{mp}\phi_{pm}T^\Lambda_m+6\Tr_m (T^\Lambda)^2_m,
\end{equation}
where
\begin{equation}
T^\Lambda_m=\sum_{q=0}^{\Lambda}\frac{1}{q+m}=\log\frac{\Lambda+m}{m}\sim\log\Lambda,  {\rm for}\ 1\leqslant m\ll\Lambda,
\end{equation}
and
\begin{equation}\label{glasses}
T^2_\Lambda=\Tr_m (T_m^\Lambda)^2=\sum_m(\sum_p\frac{1}{m+p})(\sum_q\frac{1}{m+q})\sim(2\ln^2 2+\frac{\pi^2}{6})\Lambda.
\end{equation}

The kinetic term in the matrix basis reads:
\begin{eqnarray}
&&\Delta_{mn,kl}=[\mu^2+\frac{2}{\theta}(m+n+1)]\delta_{ml}\delta_{nk}-\frac{2}{\theta}(1-\Omega^2)\\
&\times&\big[\sqrt{(m+1)(n+1)}\delta_{m+1,l}\delta_{n+1,k}+\sqrt{mn}\delta_{m-1,l}\delta_{n-1,k} \big].
\end{eqnarray}

The kinetic matrix reduces to a much simpler form when $\Omega=1$:
\begin{equation}
\Delta_{mn,kl} =\Big[\mu^2+ \frac{4}{\theta}(m{+}n{+}1)
\Big]\delta_{ml}\delta_{nk} ,
\end{equation}
which is a diagonal matrix in the double indices
and the covariance matrix $C_{mn,kl}$ reads:
\begin{equation} \label{covmatrix}
C_{mn,kl}=\frac{1}{\mu^2+ \frac{4}{\theta}(m{+}n{+}1)}\delta_{ml}\delta_{nk}.
\end{equation}
Remark that the covariance $C$ is a $\Lambda^2\times\Lambda^2$ dimensional diagonal matrix.

Since $\Omega=1$ is a fixed point of this theory \cite{MagRiv},  we shall for simplicity take
$\Omega=1$ in the rest of this paper and write the covariance as $C_{mn}=C_{mn,nm}=\frac{1}{m+n+1}$ for simplicity.

\section{The intermediate field representation and the Loop vertex expansion}
\subsection{The intermediate field representation}
The partition function for the matrix model reads:
\begin{equation}
Z(\lambda)=\int d\mu(\phi_{mn})e^{-S[\phi_{mn}]},
\end{equation}
where
\begin{equation}\label{gauphi}
d\mu(\phi_{mn})=\pi^{-N^2/2}e^{-1/2[\sum_{mn}\phi_{mn}\Delta_{mn}\phi_{nm}]}\prod_{mn} d{\rm Re}(\phi_{mm}) d {\rm Im}(\phi_{nm}).
\end{equation}
is the normalized Gaussian measure of the matrix fields $\phi_{mn}$ with covariance $C$
given by \eqref{covmatrix} and $S[\phi_{mn}]$ is the Wick ordered interaction term.

We introduce the Hermitian
matrix $\sigma_{mn}$ as an intermediate field and
the partition function can be written as:
\begin{equation}
Z(\lambda)=\int d\mu(\sigma) d\mu(\phi)e^{
-\frac{1}{2}\sum_{mnkl}\phi_{mn}\Delta_{mn,kl}\phi_{kl}-
i\sqrt{2\lambda}(\sum_{kmn}\sigma_{km}\phi_{mn}\phi_{nk}-4\sum_{mk}\sigma_{km}\delta_{mk}T^\Lambda_m)+\frac{5}{2}\lambda T^2_\Lambda}.
\end{equation}
where
\begin{equation}
d\mu(\sigma_{mn})=\pi^{-N^2/2}e^{-1/2\sum_{mn}\sigma_{mn}^2}\prod_{mn}d {\rm Re}(\sigma_{mn}) {\rm Im}(\sigma_{mn}).
\end{equation}
is the normalized Gaussian measure for the hermitian matrices $\sigma$ with the covariance:
\begin{equation}
<\sigma_{mn},\sigma_{kl}>=\int d\mu(\sigma)\sigma_{mn}\sigma_{kl}=\delta_{nk}\delta_{ml}.
\end{equation}

Then we should integrate out the original matrix fields $\phi_{mn}$, which is a $\Lambda\times\Lambda$
dimensional matrix. The result reads:
\begin{eqnarray}\label{partitionf}
&&Z(\lambda)=\int d\mu(\sigma)e^{2i\sqrt{2\lambda}\Tr T^\Lambda\sigma-
\frac{1}{2}\Tr\log[1+i\sqrt{2\lambda}\ C^{1/2}\ \hat\sigma\ C^{1/2}\ ]+\frac{5}{2}\lambda T^2_\Lambda}\nonumber\\
&=&\int d\mu(\sigma)e^{2i\sqrt{2\lambda}\Tr T^\Lambda\sigma-
\frac{1}{2}\Tr\log[1+i\sqrt{2\lambda}\ C^{1/2}(I\otimes \sigma+\sigma\otimes I)C^{1/2}\ ]+\frac{5}{2}\lambda T^2_\Lambda},
\end{eqnarray}
where $\Tr T^\Lambda\sigma=\sum_m T^\Lambda_m$, $I$ is the $\Lambda\times\Lambda$ dimensional identity matrix.
 $C^{1/2}$ is the square root of the covariance $C$ with elements:
\be
[C^{1/2}]_{mn,kl}=\sqrt{\frac{1}{m+n+1}}\delta_{ml}\otimes\delta_{nk},
\ee
the term $e^{-1/2\Tr\log[\cdots]}$ represents the $\Lambda^2\times\Lambda^2$ determinant resulting from
the Gaussian integration over $\phi_{mn}$. Here we view the vector space $R^{\Lambda^2}$ as $R^{\Lambda}\otimes R^{\Lambda}$.
For example, the operator $C^{1/2}(I\otimes \sigma+\sigma\otimes I)C^{1/2}$
transforms the basis $e_m\otimes e_n$ of $R^{\Lambda}\otimes R^{\Lambda}$ into:
\bea
C^{1/2}\ \big(e_m\otimes \sum_{k}\sigma_{nk}e_k+ \sum_{k}\sigma_{mk}e_k\otimes e_n\big)\ C^{1/2}.
\eea

However in the rest of this paper we shall simply write $\hat\sigma$ as $\sigma$ when this doesn't make confusions.
\subsection{The BKAR Tree formula and the expansion}
The most interesting quantities in our model are the
connected Schwinger's functions. Graphically connected functions
are labeled by the spanning trees.
We shall derive the connected function by the BKAR
tree formula, which is a canonical way of calculating the weight of a spanning
tree within an arbitrary graph.

Let us first of all expand the exponential as $\sum_n\frac{V(\hat\sigma)^n}{n!}$. To
compute the connected function while avoiding an additional factor
$n!$, we give a kind of $fictitious$ index $v$, $v=1,\cdots, n$ to
each field $\sigma$ in vertex $V(\sigma)$ and we could rewrite the expanded interaction term as
$\sum_n\prod_{v=1}^n \frac{V(\sigma^v)}{n!}$ . This means that we consider
$n$ different copies $\sigma_v$ of $\sigma$ with a degenerate
Gaussian measure
\begin{equation}
d\nu (\{\sigma_v\}) = d\nu( \sigma_{v_0}) \prod_{v'\neq v_0}^n  \delta(\sigma_v'- \sigma_{v_0})d\sigma_{v'} \; ,
\end{equation}
where $v_0$ is an arbitrarily marked vertex.

The vacuum Schwinger's function is given by:
\begin{theorem}[Loop Vertex Expansion \cite{Riv1}]
\begin{eqnarray}\label{LVE}
&&\log Z(\lambda, \Lambda, \cV) = \sum_{n=1}^{\infty}\frac{1}{n!}
\sum_{\cT \; {\rm with }\; n\; {\rm vertices}}   \ G_\cT
\\
&& G_\cT = \bigg\{ \prod_{\ell\in \cT} \sum_{m_\ell, n_\ell,p_\ell,q_\ell}
\big[ \int_0^1 dw_\ell \big]\bigg\} \int  d\nu_\cT (\{\sigma^v\}, \{
w \})
\nonumber \\
&& \bigg\{ \prod_{\ell\in \cT } \big[\delta_{n_\ell
p_\ell}\delta_{m_\ell q_\ell} \frac{\delta}{\delta \sigma^{v(\ell)}_{m_\ell,n_\ell}}
\frac{\delta}{\delta \sigma^{v'(\ell)}_{p_\ell,q_\ell}} \big]\bigg\}
\prod_{v=1}^n V_v \nonumber,
\end{eqnarray}
where
\begin{itemize}
\item each line $\ell$ of the tree joins two different loop
vertices $V^{v(\ell)}$ and $V^{v'(\ell)}$ which are identified through the function
$\delta_{m_\ell q_\ell'}\delta_{n_\ell
p_\ell'}$, since the propagator of $\sigma$ is
ultra-local.

\item the sum is over spanning trees joining all $n$ loop vertices. These trees have therefore
$n-1$ lines, corresponding to $\sigma$ propagators.

\item the normalized Gaussian measure $d\nu_{\cT}(\{\sigma_v\},\{w\})$ over the fields $\sigma_v$ has now a covariance
\begin{equation}
<\sigma^v _{mn}, \sigma^{v'} _{kl}>=\delta_{nk}\delta_{ml} w^\cT (v, v', \{ w\}),\\
\end{equation}
which depend on the "fictitious" indices. Here $w^\cT (v, v', \{
w\})$ equals to $1$ if $v=v'$, and equals to the infimum of the
$w_\ell$ for $\ell$ running over the unique path from $v$ to $v'$ in
$\cT$ if $v\ne v'$.
\end{itemize}
\end{theorem}

\begin{figure}[!htb]
\centering
\includegraphics[scale=0.6]{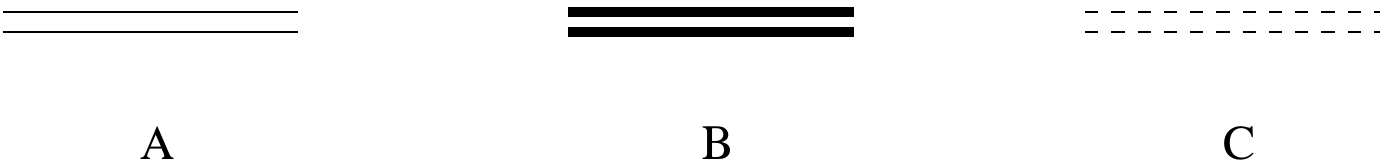}
\caption{The propagators in LVE. $A$ stands for the resolvent $R_{mn}$, B stands for the pure propagator $C_{mn}$,  and C is the propagator of the $\sigma$ fields.} \label{line}
\end{figure}

The Feynman graphs are ribbon graphs for the GW2 model \cite{Minwala,tutt,hooft,GW1}. There are three basic line structures in the LVE:
\begin{itemize}
\item The full resolvent $\hat R(\lambda, \sigma)$ is defined as:
\begin{eqnarray}\label{resol}
&&\hat R_{mn}(\sigma,\lambda)=-\frac{2}{i\sqrt{2\lambda}}\frac{\partial}{\partial\sigma_{mn}}
[-\frac{1}{2}\Tr\log(1+i\sqrt{2\lambda}C^{1/2}\sigma C^{1/2})]\nonumber\\
&=&
\big[C^{1/2}\frac{1}{1+i\sqrt{2\lambda}C^{1/2}\sigma C^{1/2}}\ C^{1/2}\big]_{mn}
\end{eqnarray}
we define the resolvent as $R=\frac{1}{1+i\sqrt{2\lambda}C^{1/2}\sigma C^{1/2}}$
and we have $\hat R=[C^{1/2}\  R\  C^{1/2}]$.

\item The propagators $C_{mn}$ between the original fields $\phi_{mn}$,
\item The propagators between the $\sigma$ fields.
\end{itemize}
The propagators are shown in Figure \ref{line}.

There are three kinds of interaction vertices in the LVE: the
counter terms, the leaf terms $K$ with coordination number $1$
and the general interaction vertices $V$. A leaf vertex $K$ is
generated by deriving once w.r.t the $\sigma$ field on the $\log_2$
term:
\begin{eqnarray}
&&K_{mn}=\frac{\partial}{\partial\sigma_{mn}}[-\frac{1}{2}
\Tr\log_2(1+i\sqrt{2\lambda}C^{1/2}\sigma C^{1/2})]\nonumber\\
&=&-\frac{1}{2}i\sqrt{2\lambda}[C^{1/2}
(\frac{1}{1+i\sqrt{2\lambda}C^{1/2}\sigma C^{1/2}}-1) C^{1/2}]_{mn}.
\end{eqnarray}

A general loop vertex could be obtained by deriving twice or more
with respect to the $\sigma$ fields:
\begin{eqnarray}\label{generalresol}
V_{m_1 m_2\cdots m_{p-1} m_p}(\lambda,\sigma)&=&\frac{\partial}{\partial\sigma_{m_1 m_2}}
\cdots\frac{\partial}{\partial\sigma_{m_{p-1} m_p}}[-\frac{1}{2}
\Tr\log(1+i\sqrt{2\lambda}C^{1/2}\sigma C^{1/2})]\nonumber\\
&=&-\frac{1}{2}(i\sqrt{2\lambda})^p(-1)^p\sum_\tau
\hat R_{m_1m_{\tau(1)}}(\sigma,\lambda)\cdots
\hat R_{m_{\tau(p)}m_1}(\sigma,\lambda).
\end{eqnarray}
with $p\ge 2$ and the sum over $\tau$ is over the $p$ cyclic permutations of the resolvents.

\begin{figure}[!htb]
\centering
\includegraphics[scale=0.5]{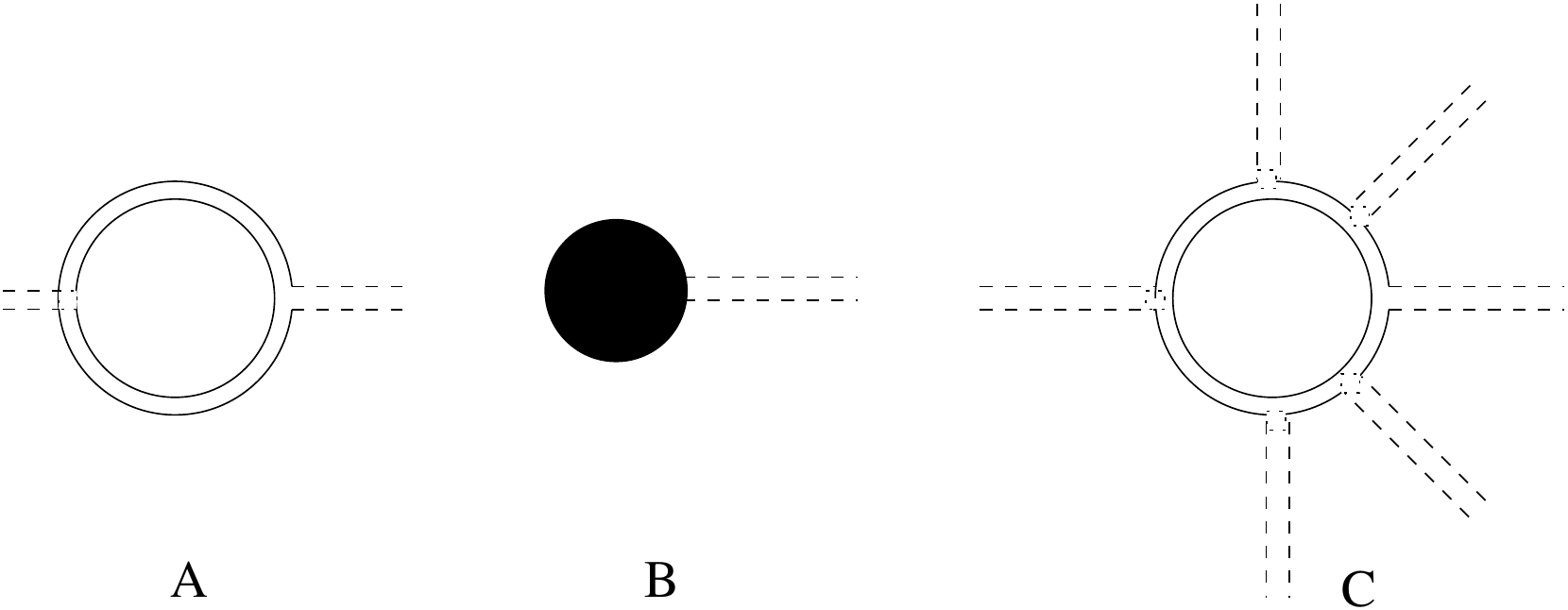}
\caption{The basic graph elements of LVE. Graph $A$ means the leaf $K$, graph $B$ means the counter term and graph $C$ means the most
general loop vertex which has several $\sigma$ fields attached. } \label{od1}
\end{figure}

The basic interaction vertices are shown in Figure \ref{od1}
where we didn't show explicitly the pure propagator $C^{1/2}$.

\subsection{The dual representation}
Since a LVE graph in the direct representation is $planar$, the notion of
duality is globally well-defined. In the dual representation we have a
canonical (up to an orientation choice) and more explicit cyclic ordering of all
ingredients occurring in the expansion (namely the resolvents, the pure propagators and the counterterms) \cite{Wanggw2,RWang}.
We shall work in the dual representation for the rest of this paper.
\begin{figure}[!htb]
\centering
\includegraphics[scale=0.6]{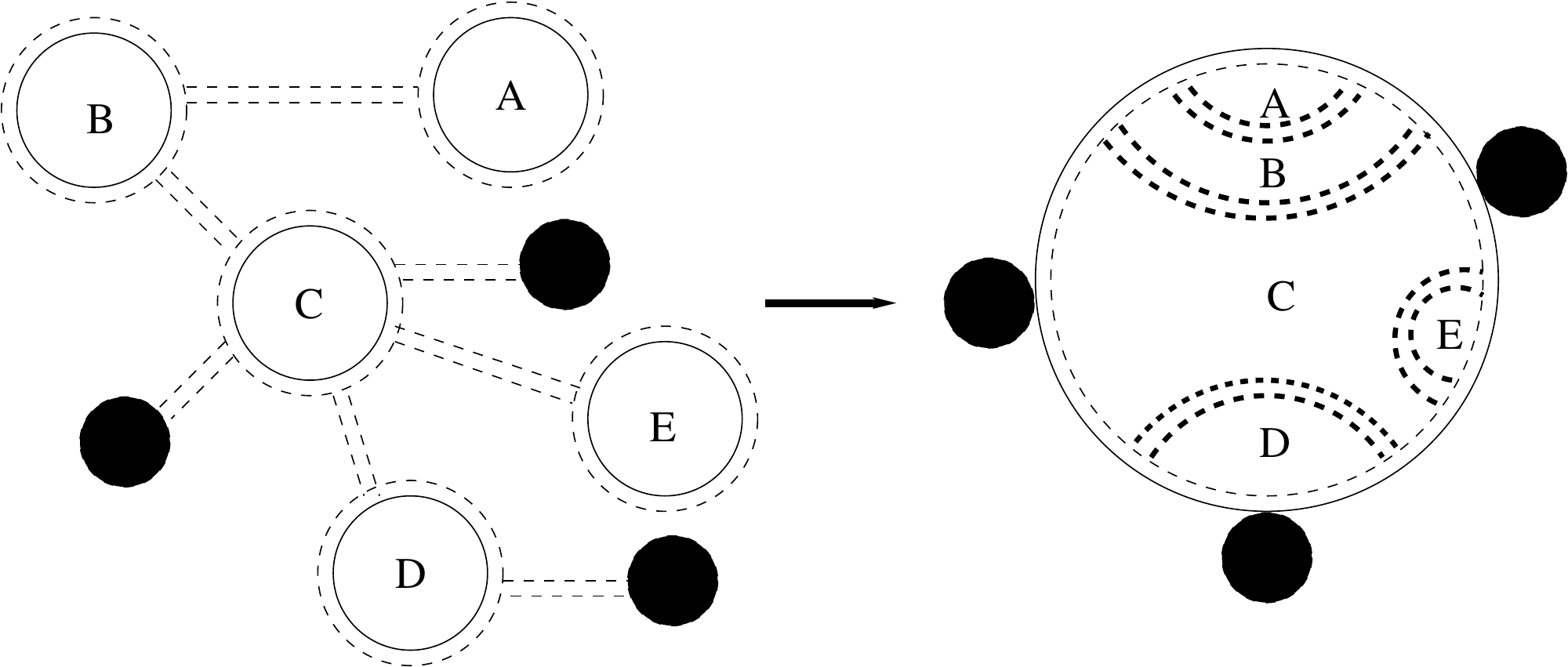}
\caption{The dual graph of a LVE. The area enclosed by the bold dash
ribbons correspond to the original loop vertices.} \label{dual1}
\end{figure}
The amplitude of an arbitrary graph with $n$ vertices and $k$ counter terms is bounded by:
\begin{equation}
G_\Lambda\le\sum_n\frac{n^{n-2}}{n!}
\lambda^n(\ln\Lambda)^{k}\le\sum (\lambda K)^n(\ln\Lambda)^k,
\end{equation}
which is divergent when $\Lambda\rightarrow\infty$. The reason is simply that the counter terms $T^\Lambda_m$
are divergent. We should introduce a renormalization process to cancel the counter terms to make the amplitude finite.
The renormalization process is called the cleaning expansion which we will introduce in the next section.

\section{The Cleaning Expansion}\label{secclean}
We shall use the multi-scale representation of the propagators and the resolvents.
Introducing the Schwinger parameter representation the propagator as:
\begin{equation}
C_{mn}=\int_0^\infty d\alpha
e^{-\alpha(\frac{\mu^2}{\theta}+m+n+1)}=K\int_0^1 d\alpha
e^{-\alpha(\frac{\mu^2}{\theta}+m+n+1)}.
\end{equation}
We decompose the propagator as:
\begin{equation}
C_{mn}=\sum_{j=0}^{\infty}C_{mn}^j,
\end{equation}
where
\begin{equation}
C_{mn}^j=\int_{M^{-2i}}^{M^{-(2i-2)}}d\alpha e^{-\alpha(\mu^2+\frac{4}{\theta}(m+n+1))},
\end{equation}
and $M$ is an arbitrary positive constant.
We could easily find that
\begin{equation}\label{sliced}
|C_{mn}^j|\le KM^{-2j}e^{-M^{-2j}||\mu^2+\frac{4}{\theta}(m+n+1)||}.
\end{equation}
We have also the sliced counter-term as:
\begin{equation}\label{fish}
T^\Lambda_m=\sum_{j=0}^{\Lambda}T^j_m.
\end{equation}

Due to the cyclic order of the global Trace operator, we could
rewrite the loop vertex in the non-symmetric from $\Tr\log[1+i\sqrt{2\lambda}\sigma C]$ and the resolvent defined in formula \eqref{resol} is written as:
\begin{equation}
R_{mn}=\frac{\partial}{\partial\sigma_{mn}}[-\frac{1}{2}\Tr\log(1+i\sqrt{2\lambda}\sigma C)]
=-i\frac{1}{2}\sqrt{2\lambda}[\frac{1}{1+i\sqrt{2\lambda}\sigma C
}]C.
\end{equation}

\begin{figure}[!htb]
\centering
\includegraphics[scale=0.6]{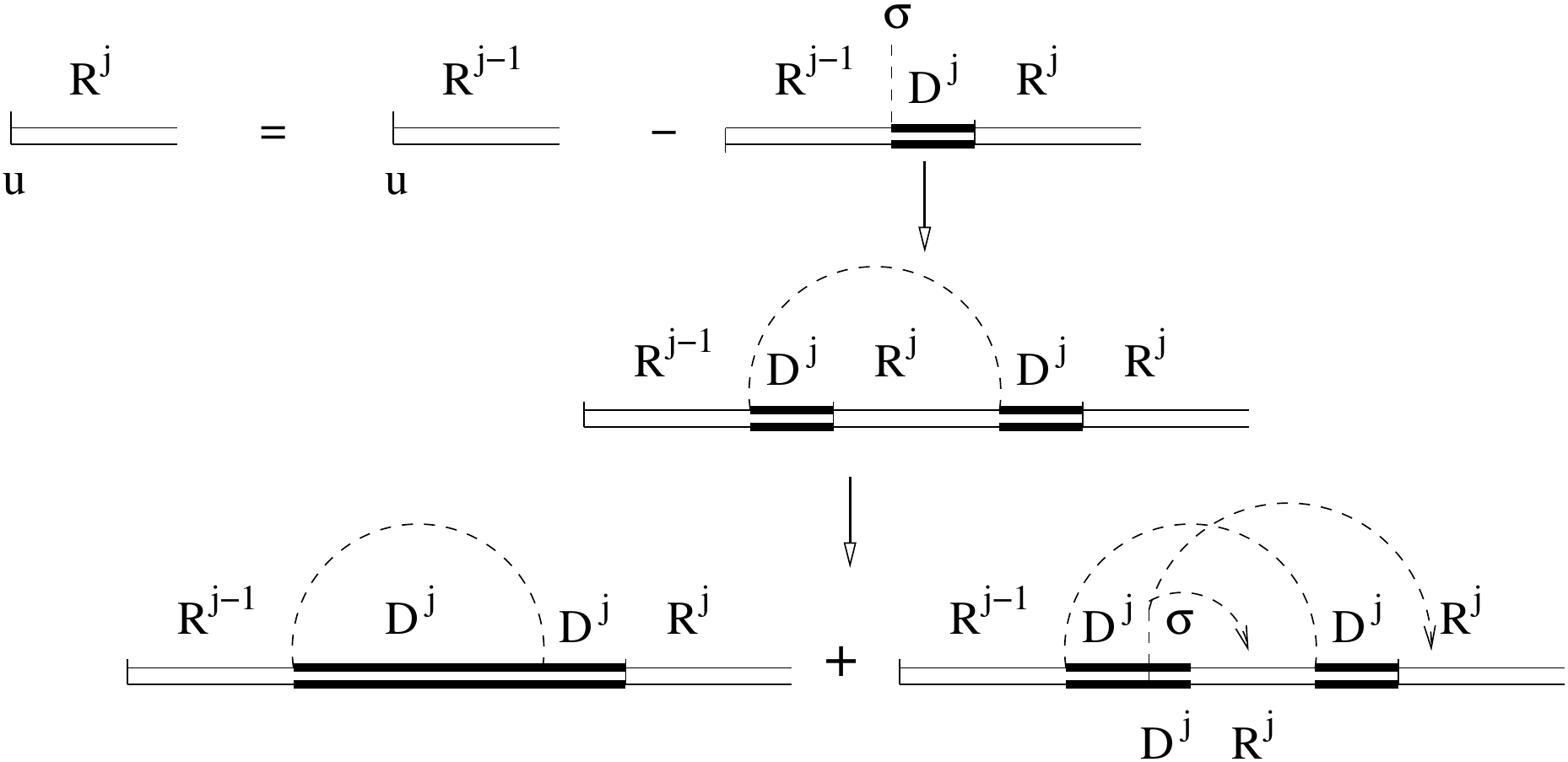}
\caption{The cleaning expansion. The ordinary ribbon stands for the resolvents and the bold double line
stands for the pure propagator. The dashed lines should be envisioned also as double lines and these stand for
the $\sigma$ fields. The L.H.S. of the last line means an inner tadpole while the R.H.S. stands for a crossing line.} \label{clean}
\end{figure}

\begin{figure}[!htb]
\centering
\includegraphics[scale=0.5]{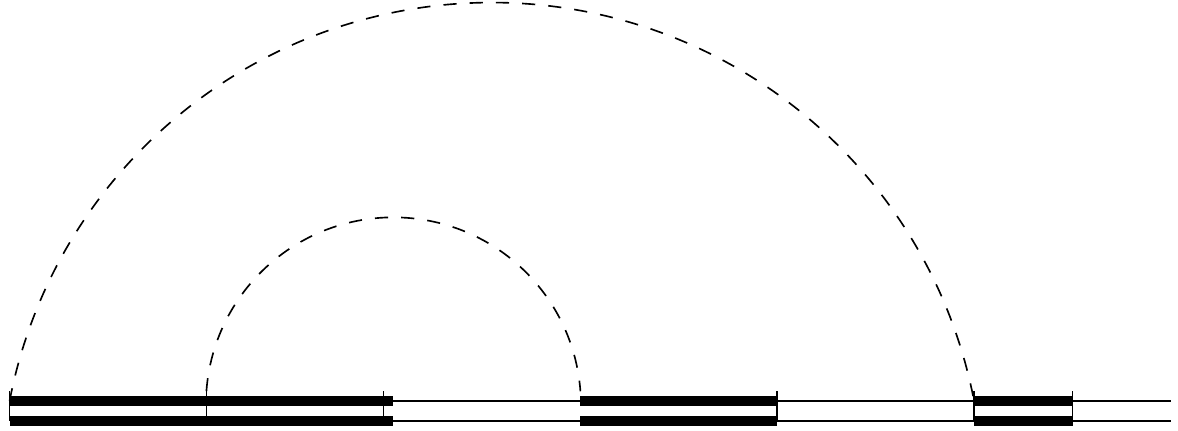}
\caption{A graph with $2$ nesting lines.} \label{nest}
\end{figure}

The idea of the cleaning expansion is to expand the contract the $\sigma$ fields hidden in the resolvent $R$
so as to compensate the counter terms and generate the convergent terms. In this process we would generate
either an inner tadpole, a crossing line (see figure \ref{clean}) or a nesting line ( see Figure\ref{nest}).
The amplitude of an inner tadpole is exactly the same as the counter term but with a minus sign,
so this compensate exactly the counter term. The amplitude of crossing or nesting line of scale $j$ is proportional to
$M^{-2j}$, so that we should generate as many of them as possible. We impose also the stopping rule for the cleaning expansion to
make sure that we don't expand forever (otherwise this would generate big combinatorial factor and make the perturbation series divergent,
for example, the number of graphs with $n$ crossings is proportional to $n!$).
We stop the expansion until we have gained enough convergent factors to compensate the divergent Nelson's factor.
A typical cleaning expansion process is shown in Figure \ref{clean}. The interested readers could look at \cite{RWang,Wanggw2}
for more details.

\section{Nelson's argument and the bound of the connected function}\label{nelson1}
In the last section we introduced the cleaning expansion, after
which all inner tadpoles should be canceled by the counter-terms in
the cleaned part of the dual graph. But there might still be
arbitrarily many counter terms in the uncleaned part and they are
divergent. Instead of canceling all of them, we re-sum them by using
the inverse formula of the Gaussian integral and integration by
parts.

We write more explicitly, but loosely, the amplitude of the connected function after the
LVE and the cleaning expansion as:
\begin{eqnarray}
A^{N}_{T}=\prod_{\bar\ell,\bar\ell_c\bar\ell_l,\bar\ell_r\in \cT}\int d\mu(w,\sigma)\int_0^1
dw_{l} \Tr\prod^\rightarrow [K^{\bar\ell}_{mn}(\sigma) R^{\bar\ell_l,\bar\ell_r}_{mn}(\sigma)
(D^{\bar\ell_c}T_\Lambda)],
\end{eqnarray}
where $R_{mn}(\sigma)$ is the full resolvent that contains the pure propagator $C$,
$\Tr $  needs to follow the cyclic order according to the real positions
of the leaf terms $K$, the resolvents $R$ and the counter terms $T_\Lambda D$. We have used the fact
that all weakening factors for the counter terms equal to one, as they are leaves in the graph.
There are only weakening factors for the $\sigma$ propagators that cross different regions in the dual graph.

Now we consider the function $G$ for which we haven't expanded the counter-term:
\begin{eqnarray}
G=\int d\nu(\sigma,w)\prod_{l\in\cT}\Tr[
K_{mn}R_{mn}(\sigma)]e^{\Tr i\sqrt{2\lambda}\sigma
T_\Lambda}.
\end{eqnarray}

We use the formula
\begin{equation}
\int d\nu(w,\sigma)f(\sigma)
g(\sigma)=e^{\frac{1}{2}\frac{\partial}{\partial\sigma}C(\sigma,\sigma',w)\frac{\partial}{\partial\sigma}}
f(\sigma) g(\sigma))|_{\sigma=0},
\end{equation}
where $C(\sigma,\sigma',w)$ is the covariance that might depend on
the weakening factor $w$ or not. Hence
\begin{eqnarray}
&&G=\int d\nu(\sigma,w)\sum_{N=0}^{\infty}\frac{1}{N!}[\frac{1}{2}\frac{\partial}{\partial\sigma}\frac{\partial}{\partial\sigma'}]^N\big\{\
\prod_{\bar l\in\bar\cT}\ \int_0^1 dw_{l'}
\Tr[K^{\bar\ell}(\sigma) R^{\bar\ell_l,\bar\ell_r}(\sigma)] e^{\Tr i\sqrt{2\lambda}\sigma T_\Lambda}]\big\}\nonumber\\
&=&\sum_{N_1=0}^{\infty}\sum_{N_2=0}^{\infty}\sum_{N_3=0}^{\infty}\prod_{l\in\cT}
\int_0^1dw_{l}\frac{1}{N_1!N_2!}(\frac{1}{2})^{N_1+N_2}[(\frac{\partial}{\partial\sigma})^{N_2}
\{\frac{\partial}{\partial\sigma}\frac{\partial}{\partial\sigma'}\}^{N_1}][K(\sigma)R(\sigma)]\nonumber\\
&&\{(\frac{\partial}{\partial\sigma'})^{N_2}
\frac{1}{N_3!}[\frac{1}{2}\frac{\partial}{\partial\sigma}\frac{\partial}{\partial\sigma'}]^{N_3}e^{\sum_q
i\sqrt{2\lambda}\sigma_{mm}
T_m^\Lambda}\}|_{\sigma=0}
.\label{nelson}
\end{eqnarray}
where we have written down explicitly the trace term in the
exponential, and the term $T_m^\Lambda$ is defined by formula
\eqref{fish}.

While the $N_1$ and  $N_2$ derivations generate connected terms, the last derivatives
generate $N_3$ disconnected terms, see graph $B$ in Figure \ref{od1}.

If we sum the the indices $m$ for the counter terms directly, we would have:
\begin{eqnarray}
&&\sum_{N_3=0}^{\infty}\frac{1}{N_3!}[\frac{1}{2}\frac{\partial}{\partial\sigma}\frac{\partial}{\partial\sigma'}]^{N_3}e^{\Tr
i\sqrt{2\lambda}\sigma T_\Lambda}|_{\sigma=0}=
\sum_{N_3=0}^{\infty} \frac{1}{N_3 !}[\frac{1}{2}\sum_{m=0}^\Lambda (i\sqrt{2\lambda} T_m^\Lambda )^2]^{N_3}\nonumber\\
&=&e^{-\lambda T^2_\Lambda}.\label{badfactor},
\end{eqnarray}
And the resumed amplitude reads:
\begin{equation}\label{naivenelson}
A^N_T=\int d\nu(\sigma, w)\prod_{\bar\ell\in\bar\cT}\
\Tr\prod^\rightarrow
[K^{\bar\ell}_{mn}(\sigma)R^{\bar\ell_l,\bar\ell_r}_{mn}(\sigma)]
e^{\Tr i\sqrt{2\lambda}\sigma T_\Lambda}e^{\lambda T^2_\Lambda},
\end{equation}

The bound in the last formula is dangerous in that $T^2_\Lambda\sim\Lambda$.
The reason for the bad factor $\Lambda$ is due to that we sum all
indices in $\sum_{m=0}^\Lambda (i\sqrt{2\lambda} T_m^\Lambda )^2$ in
formula \eqref{badfactor}, which means we detach all counter terms
in the uncleaned part of the dual graph even if they are convergent.
But this is not optimal. For example, the tadpole of amplitude $\ln(\Lambda/m)$, where $m$
is the scale of the border of the ribbon where the tadpole is
attached, is not divergent if $m\sim c\Lambda$, where $c$ is a
number that is not much smaller than $1$, say $1/2$. What's more, we
need to bear in mind that each counter term is attached to a pure
propagator in the form of $\lambda C_{nm}T^\Lambda_m$ and $C_{nm}$ decays as
$M^{-2j}$, where $2j\sim\ln\max{(m,n)}$. Here we have
ignore all the inessential factors.

So instead of considering only $T^\Lambda_m$, we need to take the
whole object $C_{nm}T^\Lambda_m$, which we also call it the full
counter term, into account. We first consider the case $m>n$, so we
have $m\sim M^{2j}$ where $M>1$ is an arbitrary constant. Then we have
\begin{eqnarray}
C_{nm}T^\Lambda_m\sim\frac{1}{m}\ln(\Lambda/m)\le\frac{\ln\Lambda}{m}.
\end{eqnarray}
So that the counter term doesn't cause any
divergence as long as $m>\ln\Lambda$ and we could just bound these
counter terms by $1$ and need not to detach them from the connected
graph. The counter terms become dangerous only when $m\le\ln\Lambda$
and we need to re-sum them as introduced before.

So that we only need to sum over the indices $m$ for $m\le\ln\Lambda$ in formula \eqref{badfactor},
which reads:
\begin{eqnarray}
&-&2\lambda\sum_{m=1}^{\ln\Lambda}\ln^2(\Lambda/m)<
-2\lambda\sum_{m=1}^{\ln\Lambda}\ln^2\Lambda=-2\lambda(\ln\Lambda)^3,
\end{eqnarray}
and the amplitude of the connected function after the resummation reads:
\begin{equation}
A^N_T=\int d\nu(\sigma, w)\prod_{\bar\ell\in\bar\cT}\
\Tr\prod^\rightarrow
[K^{\bar\ell}_{mn}(\sigma)R^{\bar\ell_l,\bar\ell_r}_{mn}(\sigma)]
e^{\Tr i\sqrt{2\lambda}\sigma T_\Lambda}e^{\lambda (\ln\Lambda)^3},
\end{equation}

\begin{figure}[!htb]
\centering
\includegraphics[scale=0.6]{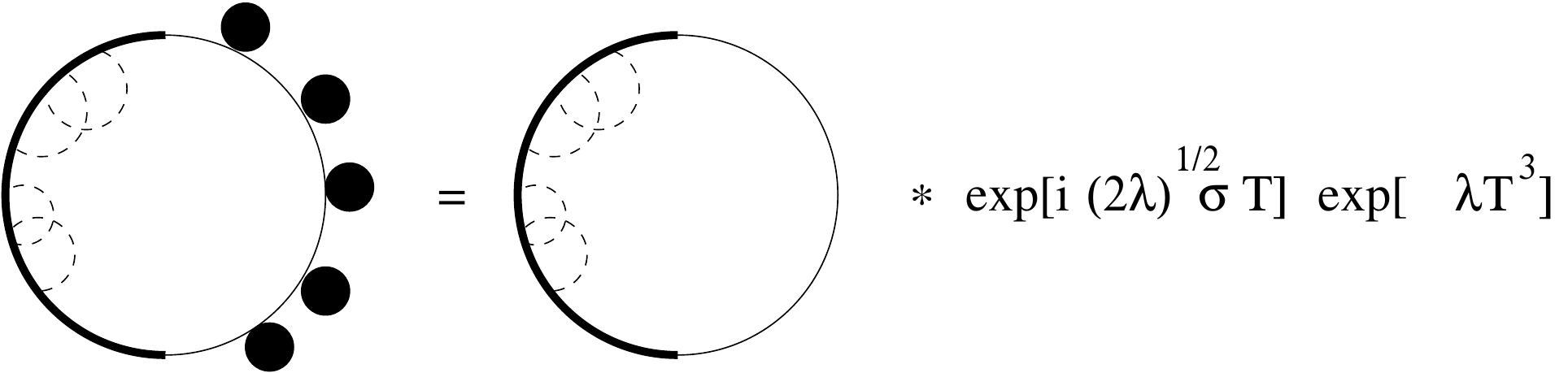}
\caption{A sketch of the resummation of the counter terms. Where the term $T$ means $\ln \Lambda$. The counterterms with the scale index of outer border
larger than $\ln\ln \Lambda$ are not summed.} \label{resumm}
\end{figure}

This resummation process is shown in Figure \ref{resumm}.

The divergent factor $e^{\lambda(\ln\Lambda)^3}$ is not dangerous as we
could bound it by the following formula:
\begin{equation}
e^{-a j^3_{max}}\cdot (j^2_{max})!\cdot e^{\lambda \ln^3\Lambda}\sim
e^{-a j^3_{max}+ 2j^2_{max}\ln  j_{max}+\lambda  j^3_{max}}<1
\end{equation}
as long as $a$ is chosen properly, for example  $a=1.4\lambda$.
Here we have use the fact that fact that $\ln\Lambda\sim j_{max}$ and
$(aj^2_{max})!\sim e^{2j^2_{max}\ln j_{max} }$.
\section{The Borel summability}
In this section we consider the Borel summability of the perturbation series \cite{Riv}.
\begin{theorem}
The perturbation series of the connected function for $\phi^4_2$ theory is Borel summable.
\end{theorem}
\prf
For the perturbation series $\sum_{n=0}a_k\lambda^k$  to be Borel summable to
the function $G$, we need to have
\begin{equation}\label{borelexp}
G(\lambda)=\sum_{n=0}a_k\lambda^k+R^{(n+1)}(\lambda),
\end{equation}
where $R^{(n+1)}(\lambda)$ is the Taylor remainder. The analyticity domain $C_\lambda$ for $G$
should be at least $|\lambda|<\frac{1}{K}$ and ${\rm Re}\lambda>0$ \cite{Riv}, which means
\begin{equation}-\frac{\pi
}{4}\le{\rm Arg} \sqrt \lambda\le \frac{\pi}{4}.
\end{equation}

We rewrite the resolvent as
\begin{equation}
R=\frac{1}{1+i\sqrt{2\lambda} C^{1/2}\sigma C^{1/2}}.
\end{equation}
Since the matrix $C^{1/2}\sigma C^{1/2}$ is Hermitian, its eigenvalues are real, so that
there are no poles in the denominator.
In the analytic domain of $G$ we have
\begin{equation}
\parallel R\parallel=|\frac{1}{1+i\sqrt{2\lambda} C^{1/2}\sigma C^{1/2}}|\leqslant\sqrt{2},
\end{equation}
 and
\begin{equation}
\parallel K\parallel=\parallel R-1\parallel\leqslant 1+\sqrt{2}.
\end{equation}

However in the analytic domain $C_\lambda$ the linear counter term becomes:
\begin{equation}\label{analytic1}
e^{\Tr i\sqrt{2\lambda}\sigma T_\Lambda}=e^{\Tr i|\sqrt{2\lambda}|\cos\theta\sigma T_\Lambda}e^{-\Tr|\sqrt{2\lambda}|\sin\theta\sigma T_\Lambda},
\end{equation}
where $\theta={\rm Arg} \sqrt \lambda$.
We could bound the first term in \eqref{analytic1} by $1$, but the second term would diverge for negative $\sigma$.

We rewrite this term as:
\begin{eqnarray}
\int d\mu(\sigma)e^{-1/2\Tr(\sigma+\sqrt{2\lambda}\sin\theta T_\Lambda)^2}e^{\sin^2\theta T^2_\Lambda}.
\end{eqnarray}
The term $e^{\sin^2\theta T^2_\Lambda}$ could diverge at worst as $e^{1/2 T^2_\Lambda}$ for $\theta=\pm \pi/4$.
But this is not dangerous, since we could still bound it with the convergent factor $e^{-aj^3_{max}}\sim e^{-aT^3_\Lambda}$ that we gained
from the crossings and nesting lines.

We use simply the Taylor expansion with remainder for the connected function \eqref{borelexp}:
\begin{equation}
G(t\lambda)|_{t=1}=\frac{G^{(n)}(\lambda)}{n!}|_{t=0}+\int_0^1dt\frac{(1-t)^n}{n!} G^{(n+1)}(t\lambda),
\end{equation}
followed by explicit Wick contractions. We have for the reaminder
\begin{equation}
||R^{n+1}||<|\lambda|^{n+1}K^n(2n)!!\le|\lambda|^{n+1}[K']^n(n!),
\end{equation}
where $K$ and $K'$ are positive numbers including the possible factors $\sqrt{2}$ or $\sqrt{2}+1$ from the
bound of the resolvent $R_{mn}$ and of the leaf $K_{mn}$ respectively.
Hence we have proved the Borel summability of the perturbation series. \qed

\section{Conclusions and prospectives}
In this paper we have constructed the $2$-dimensional Grosse Wulkenhaar model with the method
of loop vertex expansion (LVE). The next step should be constructing the $4$ dimensional case, which is
the real interesting one. In this case we need also to consider the 4-point function and the abstract cluster
expansion would play a more important role. This work is still in progress.

\end{document}